# Enhanced optical Kerr nonlinearity of graphene/Si hybrid waveguide

Qi Feng[1], Hui Cong[1], Bin Zhang[1,2], Wenqi Wei[1,2,3], Yueyin Liang[1,2], Shaobo Fang[1], Ting Wang[1*] and Jianjun Zhang[1]

[1]Institute of Physics, Chinese Academy of Sciences, Beijing, 100190, People's Republic of China
[2]School of Physical Sciences, University of Chinese Academy of Sciences, Beijing, China, 100190
[3]School of Physics and Technology, Wuhan University, Wuhan, 430072, China

**Abstract**
In this work, we experimentally study the optical kerr nonlinearities of graphene/Si hybrid waveguides with enhanced self-phase modulation. In the case of CMOS compatible materials for nonlinear optical signal processing, Si and silicon nitride waveguides have been extensively investigated over the past decade. However, Si waveguides exhibit strong two-photon absorption (TPA) at telecommunication wavelengths, which lead to a significant reduction of nonlinear figure of merit. In contrast, silicon nitride based material system usually suppress the TPA, but simultaneously leads to the reduction of the Kerr nonlinearity by two orders of magnitude. Here, we introduce a graphene/Si hybrid waveguide, which remain the optical properties and CMOS compatibility of Si waveguides, while enhance the Kerr nonlinearity by transferring patterned graphene over the top of the waveguides. The graphene/Si waveguides are measured with a nonlinear parameter of 510 $W^{-1}m^{-1}$. Enhanced nonlinear figure-of-merit (FOM) of 2.48 has been achieved, which is three times higher than that of the Si waveguide. This work reveals the potential application of graphene/Si hybrid photonic waveguides with high Kerr nonlinearity and FOM for nonlinear all-optical signal processing.

**Introduction**
Optical nonlinear effects in CMOS-compatible integrated optical devices are of great significance as they can be explored to realize a variety of functionalities ranging from all-optical signal processing to light generation [1-5]. Silicon-on-insulator (SOI) has been regarded as a popular platform for ultra-dense on-chip integration of photonic and electronic circuitry due to its compatibility with CMOS fabrication. In addition, nonlinear optical properties of silicon waveguides are also heavily explored over the past decade, such as stimulated Raman scattering (SRS), Raman amplification, self-phase modulation (SPM), four-wave mixing, super-continuum generation [2,6-8]. However, the existence of two-photon-absorption (TPA) at telecom wavelength (around 1550 nm) in Si platform leads to a strong degradation in the value of nonlinear figure-of-merit, which limits the power efficiency of nonlinear functionalities. In addition, TPA increases the photon loss in the process as well as generating carriers subsequently producing usually undesired free-carrier absorption (FCA) and free-carrier dispersion (FCD). TPA and FCA generally cause optical losses, which in turn lower the peak power inside the waveguide and therefore reduce the conversion efficiency of optical nonlinear process [9,10]. There are other promising platforms such as chalcogenide glass and AlGaAs, which possess high nonlinearity and low TPA, but highly challenging fabrication processes limit their usage in CMOS compatible applications [11-16].

Furthermore, CMOS compatible platforms such as $Si_3N_4$ and Hydex glass exhibit low TPA at telecom wavelength, thus efficiently reducing the nonlinear loss as well as linear loss, however, their nonlinear refractive index is approximately one order of magnitude smaller than that of silicon [17-22]. Therefore, the best way to fulfill the requirement of silicon nonlinear photonics is to integrate them with novel materials with high Kerr coefficient while keeping the silicon platform for its economic advantages. One of the best candidate is graphene, which has outstanding optoelectronic properties, while remians compatibilitiy for integration with silicon photonic devices. In the field of photonics, graphene has enjoyed widespread research attention in various optical devices, such as photodetectors, modulators, optical switches, optical gates, and lasers [23-29]. Some of the strengths of graphene manifest in its unique optical properties, the most common include its large tunable refractive index, high confinement factor, and a universal absorption of 0.3 %. One of the major optical properties of graphene is the giant optical Kerr nonlinearity which has been previously reported by several groups with Kerr-coefficients ranging from $10^{-7}$ to $10^{-13}$ $m^2/W$ [30-35]. In this paper, we demonstrate experimental studies of SPM process under picosecond pulses in Si waveguides and graphene/Si (G/Si) hybrid waveguides. The positioning of graphene on Si waveguides was done by precise transfer process. The effective Kerr coefficient $n_2$ of graphene/Si hybrid waveguide is calculated to be~ $2\times10^{-17}$ $m^2/W$, which is three times higher than that of Si waveguide. Furthermore, the FOM has been enhanced from 0.7 in Si waveguides to 2.48 in G/Si hybrid waveguides.

**Design and Fabrication**

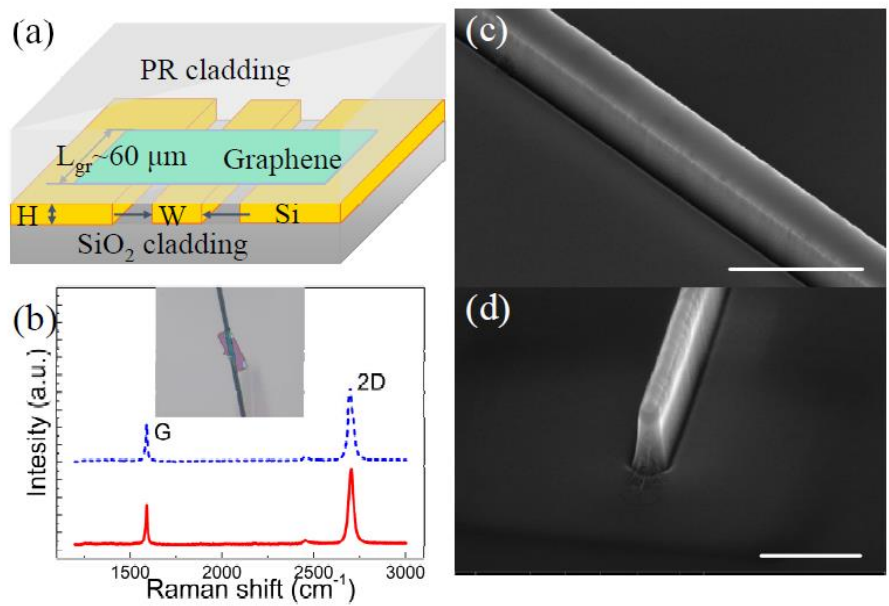

Fig. 1: (a) Schematic diagram of the graphene-Si hybrid waveguide. W, H., $L_{gr}$ denote the width and height of Si wavegudie, and length of graphene covering on Si waveguide, respectively. Raman spectra (b) of the single layer graphene before (red solid curve) and after (blue dotted curve) transferring onto Si waveguide. Inset: optical microscope image of graphene on Si waveguide. Tilted SEM image of (c) sidewall and (d) inverse taper of Si waveguide, with scale bar of 1 μm and 500 nm, respectively.

Here, we use single-mode silicon waveguides with dimensions of 500 nm × 340 nm × 3.5 mm, fabricated on a SOI wafer with a buried oxide layer of 3 μm. Standard e-beam lithography is used to pattern Si waveguides in JEOL-6300 (100kV) system with ZEP-520A EB resist. After development, the pattern is transferred by inductively coupled plasma (ICP) etching in Oxford PlasmaPro 100. The inverse taper with tip width of 100 nm and tip length of 50 μm are designed at each end of the waveguides to improve the efficiency of adiabatic coupling. Chemical vapor deposition (CVD)-grown monolayer graphene with a length of 60 μm was then transferred onto the waveguides by precise positioning [36]. The schematic diagram of the G/Si hybrid waveguide is shown in Figure 1(a), while (c) and (d) show good quality of sidewall of Si waveguide and inverse taper coupler.

The explicit characterization of monolayer graphene is obtained by Raman spectroscopic measurements, pumped with a 488 nm laser. In Figure 1(b), Raman spectrum of the graphene layer coated waveguide device is measured by LabRAM HR 800 spectrometer. Both of the spectra show a G peak (~ 1586 $cm^{-1}$) with a full width at half maximum (FWHM) of ~ 18 $cm^{-1}$ and a 2D peak (~ 2700 $cm^{-1}$), with a 2D-to-G peak intensity ratio of about 1.2, implying that the transferred graphene is monolayer and the corresponding chemical potential is around 0.2 eV. Raman spectra of graphene on different waveguides are measured for verification of graphene transfer reliability [34]. The results turn out to be similar as the Raman spectra presented in Figure 1(b).

For comparison, identical silicon waveguides without graphene were fabricated under the same procedure. In order to investigate the third-order optical nonlinearity of G/Si hybrid waveguides, the dispersion and group velocity dispersion (GVD) are calculated using the standard Sellmeier equation by Lumerical MODE solutions, as shown in Figure 2. The group velocity dispersion is calculated to be -4.5 $ps^2$m at the wavelength of 1550 nm. It is noted that comparing with the Si waveguide, the graphene has negligible effect on the GVD.

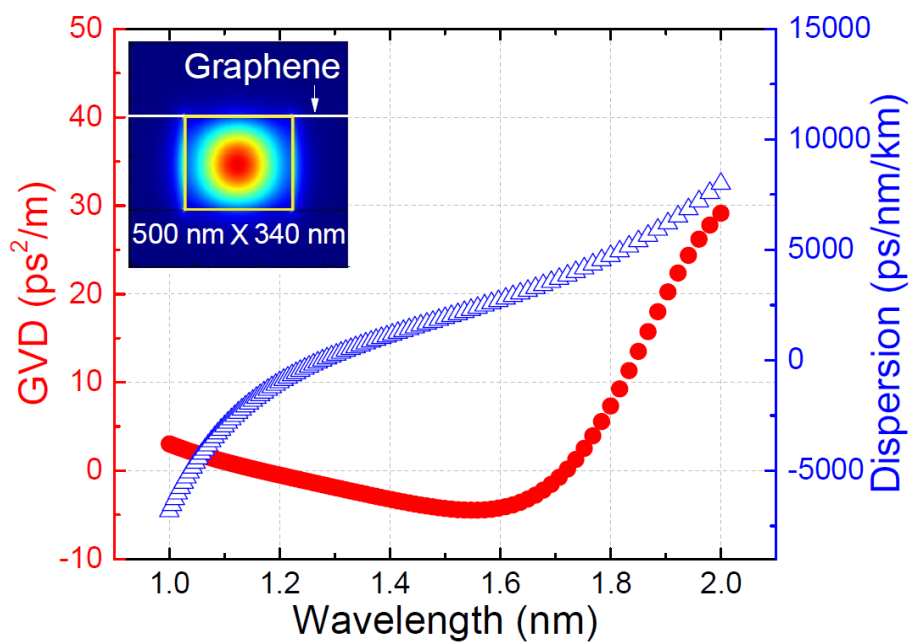

Fig.2 : GVD and dispersion as a function of wavelength for G/Si hybrid waveguide. Inset: fundamental TE-mode optical intensity distribution of the G/Si hybrid waveguide, which were used for nonlinear refractive index calculation. Graphene layer is indicated as solid white line, and silicon core is indicated inside the surrounded by yellow lines.

**Results and Discussion**

Both Si and G/Si waveguides are pumped by a PriTel's FFL series of picosecond fiber laser. The laser produces pulses with a center wavelength of around 1548 nm at a repetition rate of 20 MHz, and a pulse duration of 1.5 ps, which are delivered with a polarization-maintaining (PM) fiber. The pulses were coupled into and out of the waveguides devices via lensed fibers and inverse taper mode-converter with a coupling loss of approximately 10 dB per facet. It is noted that one meter long PM fiber between the pulse laser and the waveguide devices was chosen in order to eliminate spectrum change induced from the nonlinear effects within the fiber. To monitor the input spectrum, a 90:10 coupler is inserted in the setup to split off 10 % of the laser power into optical spectrum analyzer (OSA, Yokogawa AQ6370D). The propagation loss measurements were carried out on Si waveguides by cut-back method. And the linear propagation loss in silicon waveguide is estimated to be (3.5±0.5) dB/cm. Here TE polarized light was coupled into the waveguide by PM fiber after polarization-controller, and the output power was monitored by fiber optic power meters (Thorlabs PM20).

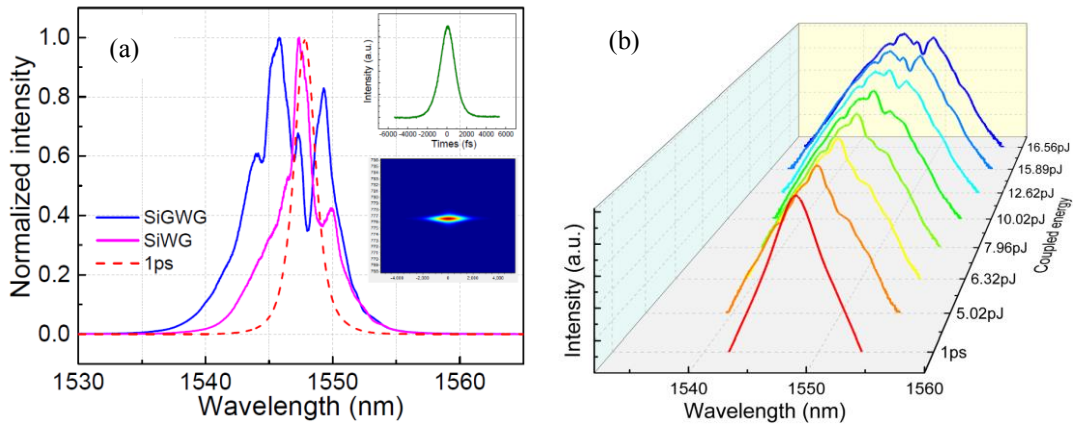

Fig.3: Experimental results of the transmission spectra of (a) comparison between the Si (magenta solid curve) and G/Si hybrid (blue solid curve) waveguides under the same input energy with 1.5 ps input pulse (spectrum denoted as red dashed curve). Inset: autocorrelation spectrum in time domain. (b) the output spectra of G/Si hybrid waveguide under various input energies. The red curve denotes the 1.5 ps pulse spectrum.

Prior to the SPM measurements, as shown in inset of Figure 3(a), we measure the temporal characteristics of the input pulse by means of frequency-resolved optical gating (FROG) instrument (Coherent Solutions HR150). The Gaussian-shape pulse with a pulse width of 1.5 ps has been confirmed. The actual SPM measurements consist of simultaneously recording the spectral widths of the input and output spectra for the waveguides with graphene lengths of 60 μm. Here, the measured propagation loss induced from graphene absorption is 0.045±0.010 dB/μm.

SPM measurements were carried out by measuring the transmission spectra in both Si waveguides and G/Si hybrid waveguides. It is shown in Figure 3(a) that the dotted red curve represents the spectrum of original input pulse, while the green and blue solid curves represent the output spectra for both Si waveguide and G/Si waveguide under identical input pulse energy, respectively. To note, both waveguides have the same length of 3.5 mm. The input pulse has spectral linewidth of 2.1 nm as shown in the dash curve of Figure 3(a). By comparing spectrum of Si waveguide with input pulse, the spectral broadening is assumed to be significantly less than the G/Si hybrid waveguides, which exhibits a strong $1.8\pi$ phase change. This result leads to the strong enhancement of third-order nonlinearity by introducing graphene decoration over the top of Si waveguides. The energy dependent measurements of G/Si hybrid waveguide are shown in Figure 3(b), with input pulase energies ranging from 5 pJ to 16 pJ. Since SPM alone is known to yield a symmetric spectral distribution around the injected laser frequency, the asymmetry must stem from other factors such as chirped injected laser pulses, self-steepening, GVD or changes in the free carrier density by TPA. Self-steepening can be ruled out. First, the change of $n_2$ within narrow spectral bandwidth of the ps-pulse is negligible. Secondly, self-steepening induced spectral red-shift is absent in our experiments [37,38]. GVD should be of minor influence as well, as will be discussed in detail later. Therefore, FCA and TPA effects play dominant role in our case.

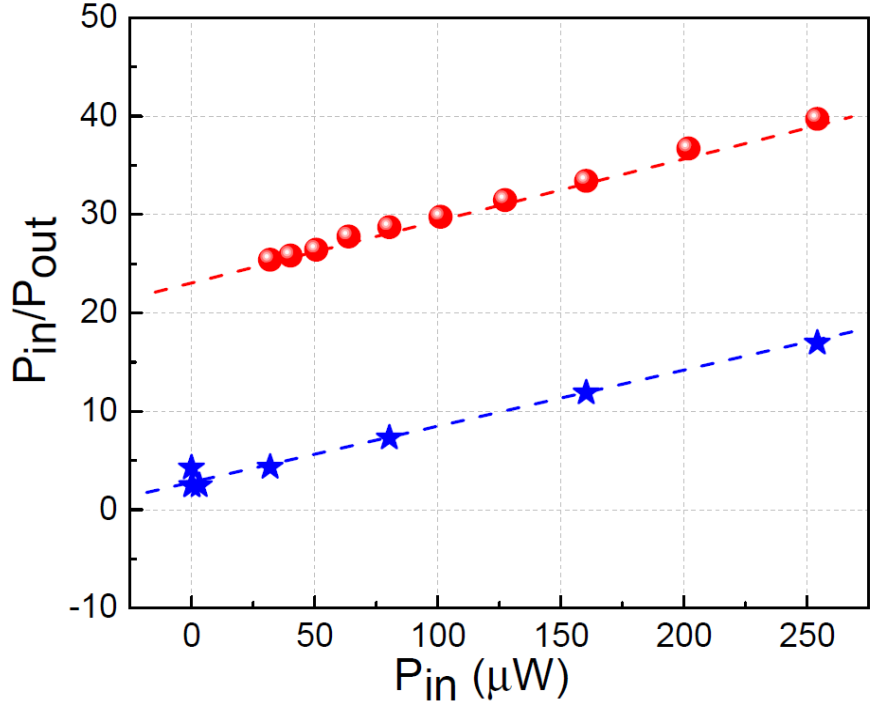

Fig.4: Pin/Pout vs Pin for the Si waveguide and G/Si hybrid waveguide, denoted as blue stars and red circles, respectively. Linear fit for both Si waveguide and G/Si hybrid are denoted as dashed lines.

Prior to the analysis of the nonlinear transmission, the relation between the average output power $P_{out}$ and the average input power $P_{in}$ is recalled in case of a dominant TPA effect:

$$\frac{P_{in}}{P_{out}} = 2Im(\gamma)L_{eff}e^{\alpha L}P_{in} + e^{\alpha L} \qquad (1)$$

where α=3.5 dB/cm is the linear propagation loss, $Im(\gamma)=\beta_{TPA}/(2A_{eff})$ is the imaginary part of the γ nonlinear coefficient due to TPA, L is the waveguide length and $L_{eff}$ the effective optical path length reduced by the linear propagation loss through $L_{eff}$=(1-$e^{-\alpha L}$)/α.

Thus, this equation discloses a linear relation between the ratio $P_{in}$/$P_{out}$ and the measured input power $P_{in}$ with the slope being proportional to the nonlinear coefficient $\beta_{TPA}$. From the similar slope shown in Figure

4, the TPA coefficients $\beta_{TPA}$ for both Si waveguide and G/Si hybrid waveguide can be extracted with similar values of 0.5 cm/GW.

According to this result, it is verified that the effective Kerr nonlinearity is significantly larger for G/Si hybrid waveguides. The output spectra from G/Si hybrid waveguides are depicted in Figure 3(b). It shows that with increasing $P_{in}$, the optical Kerr effect induced self-phase modulation becomes asymmetric, which will be discussed in details as below.

**NLSE simulation**

We calculated the case where a Gaussian-shape laser pulse is coupled into the Si waveguide and G/Si hybrid waveguide. By using the nonlinear Schrodinger equation (NLSE) with the split-step Fourier method, the on-chip SPM process can be simulated with the equation below [37]

$$\frac{\partial A}{\partial z} = -\frac{1}{2}\alpha A + i\sum_{m=2}^{10}\frac{i^m \beta_m}{m!}\frac{\partial^m A}{\partial \tau^m} + i\gamma|A|^2 A - \frac{\sigma}{2}(1+i\mu)N_c A \quad (2)$$

where A(z,t) is the slowly varying temporal envelop along the length z of a nonlinear medium, $\gamma_0=\omega_0 n_2/cA_{eff}$ is the nonlinear parameter, $\omega_0$ is the optical frequency, $\beta_m$ is the m-th order dispersion coefficient, $n_2$ is nonlinear Kerr coefficient, $\beta_{TPA}$ is the two-photon absorption coefficient, $A_{eff}$ is the effective mode area, $N_c$ is the free carrier density, $\sigma$ is the free carrier absorptin coefficient, $\mu$ is the free carrier dispersion coefficient, $\alpha$ is the linear loss parameter. Thus $N_c$ can be obtained by solving:

$$\frac{\partial N_c}{\partial t} = \frac{\beta_{TPA}}{2\hbar\omega}\frac{|A|^4}{A_{eff}^2} - \frac{N_c}{\tau_c} \quad (3)$$

where $\tau_c$ is the effective lifetime of free carriers with an estimated value of 0.5 ns. Therefore, the profile of $N_c(t)$ is calculated by solving Equation (3) approximately near the front end of the waveguide where $|E(z,t)|^2$ is close to its input. Noting that pulse width $T_0<\tau_c$, the $\tau_c$ term can be ignored as carriers do not have enough time to recombine over the pulse duration. The carrier density is then given by:

$$N_c(t) \approx \frac{\beta_{TPA}I_0^2 T_0}{2h\nu_0}\sqrt{\frac{\pi}{8}}[1+\text{erf}(\frac{\sqrt{2}t}{T_0})] \quad (4)$$

The pulse dynamics are governed by the interplay of SPM and dispersion whose relative strengths can be determined by several characteristic lengths, namely the GVD and third-order dispersion (TOD) lengths, defined as $L_D = T_0^2/|\beta_2|$ and $L'_D = T_0^3/|\beta_3|$, respectively, and the nonlinear length, defined as $L_{NL} = 1/\gamma P_0$.

Beside TPA, free-carrier effect within silicon waveguides, such as FCD and FCA, could lead to the asymmetry of SPM spectrum of G/Si hybrid waveguides. Especially in the case of FCD effect, it can lead to the decrement of refractive index and thus cause the acceleration of the pulse. The last term of Equation 2 reveals that free carriers interact with the optical pulse by both modulating its phase through FCD (which acts couter to the Kerr effect), as well as reducing intensity by FCA. As depicted by the rate equation (Equation 3), the generation of free carriers in time follows the evolution of the pulse intensity squared – that is, the free-carrier concentration will be negligible at the leading edge of the pulse and be significant at the trailing edge. Therefore, FCA causes nonliear absorption of the trailing edge of the pulse, generating pulse asymmetry. In the case of G/Si hybrid waveguides, the enhanced free carrier dentiy $N_c$ introduced by graphene could lead to a strong spectral asymmetry in the SPM measurement. Here, the carrier density of $5.85\times10^{16}$ cm$^{-3}$ and $8.15\times10^{16}$ cm$^{-3}$, the free carrier absorption coefficient of

$1.45×10^{-17}$ cm$^{-2}$ and $6×10^{-17}$ cm$^{-2}$, for Si and G/Si hybrid waveguides, respectively, were applied to our numerical calculation.

Here, the nonlinear length $L_{NL}$, dispersion length $L_D$, are calculated to be 0.65 mm and 19.26 mm, respectively. Given that the $L_D$ is much longer and the $L_{NL}$ is much shorter than the waveguide length (3.5 mm) for both Si and G/Si hybrid waveguides, the pulse dynamics will be dominated by the third-order nonlinearity rather than the dispersion in the wavegudies.

Equations (2) and (3) are then solved using a split-step Fourier transform method to model the behaviour of the pulse propagation in Si and G/Si hybrid waveguides. As shown in Figure 5, the simulated spectra have relatively good agreement with our experimental results under various input energies from 5-16 pJ. The extracted $\overline{n_2}$ value of G/Si hybrid waveguide is here calculated to be $2×10^{-17}$ m$^2$/W, which is three times larger than that of the Si waveguide.

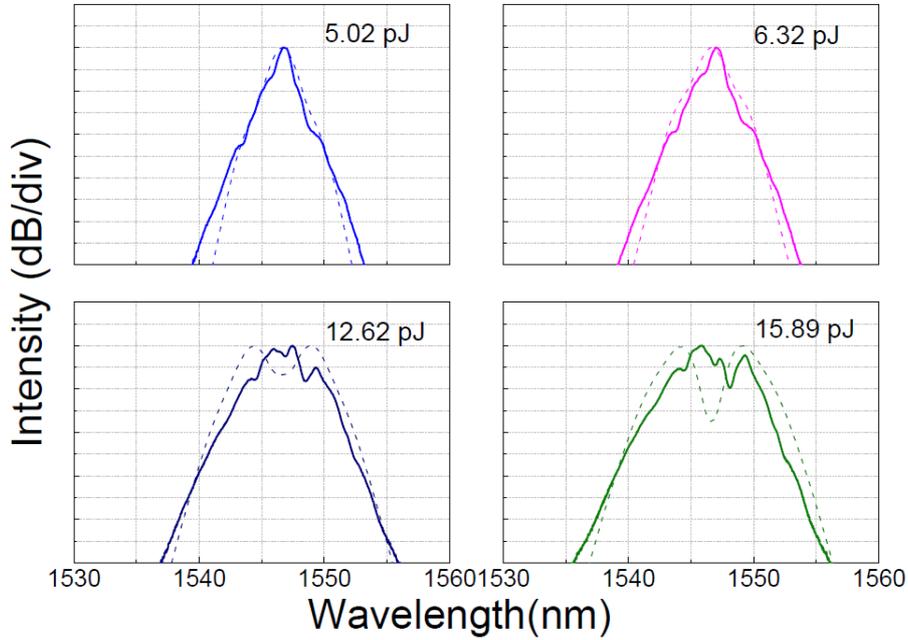

Fig.5: Experimentally measured and numerically calculated spectra of the output picosecond pulse propagating along G/Si hybrid waveguide for various coupled energy, denoted as solid and dashed curves, respectively.

As shown in Figure 6, we compared the spectral broadening for Si waveguide and G/Si hybrid waveguide as a function of coupled peak power. It is shown that spectral broadening of 3.5 mm long Si and G/Si hybrid waveguides are around 11.5 nm and 14.7 nm, respectively. The peak phase shift $\Phi_{max}$ (in radians) experienced by the pulse is given by $\Phi_{max} = 2\pi \frac{n_2 P L}{\lambda A_{eff}}$, where P is the peak pulse power. Using the effective area $A_{eff}$ of 0.16 $\mu$m for both Si and G/Si waveguides, the maximum phase shift $\Phi_{max}$ are extracted as 0.54π and 1.8 π, respectively. The enhanced Kerr nonlinearity in G/Si waveguide results in additional 1.3π phase shift with calculated nonlinear parameter γ of 510 W$^{-1}$/m.

Nonlinear figure-of-merit can be defined as FOM= $n_2/(\lambda\beta_{TPA})$, which is a measure of the optical nonlinear efficiency of the medium when both the nonlinear refractive index and nonlinear loss mechanisms are accounted for [39,40]. In addition, it provides a useful dimensionless measurement of suitability of the material for nonlinear switching. For a nonlinear directional coupler, the required nonlinear phase shift for optical switching is $4\pi$ and thus it must satify FOM > 2 for such devices [41]. For other devices, such as a nonlinear Mach-Zehnder interferometer, a $\pi$ phase change is sufficient and the nonlinear FOM only needs to satisfy the condition FOM > 1/2. Silicon platform normally exhibit nonlinear FOM of ~0.6 at 1.55 $\mu$m, which is insufficient for optical switching applications. In this work, the FOM of G/Si hybrid waveguide is calculated to be approximately 2.48, which is higher than that of Si, SiGe and hydrogenated amorphous-Si waveguides, as shown in Table 1. In addition, although chalcogenide glass and AlGaAs possess high nonlinearity and low TPA, they are limited to applications where CMOS compatibility is not required due to the challenging fabrication for highly efficient waveguides. While for other platforms such as silicon nitride and Hydex glass, they are COMS compatible and exhibit much lower nonlinear loss and linear loss due to low TPA at telecom wavelength, however, the refractive index is ten times smaller than that of silicon. A key goal of all-optical chips is to reduce both the device footprint and the operation power. The significant improvement in both the optical kerr nonlinearity and nonlinear FOM in G/Si raise the prospect to provide a truly practical and viable platform for nonlinear photonic applications in the telecommunication wavelength window. This reveals that the corporation of single layer graphene can be employed to increase the nonlienar performance of silicon-based waveguides in all-optical signal processing.

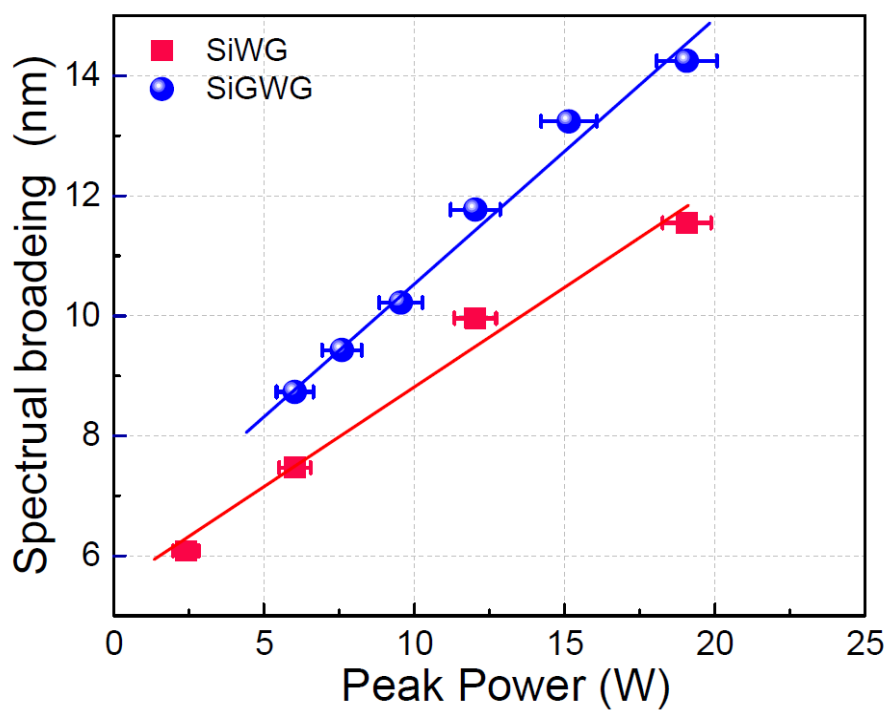

Fig.6: Measured SPM-induced spectral broadening for both Si and G/Si hybrid waveguides, denoted as red squares and blue circles, respectively, as a function of the coupled peak power.

**Conclusion**

In conclusion, enhancement of third-order nonlinearity in G/Si hybrid waveguide has been studied here by self-phase modulation experiments, and enhanced spectrum broadening has been observed in G/Si hybrid waveguide. Although the decorated graphene exhibits a relatively weak evanescent fields in such structure, three times magnitude enhanced Kerr nonlinearity is still achieved on G/Si hybrid waveguide with an overall optical nonlinear parameter of 510 W$^{-1}$/m. The FOM has been improved as well from 0.7 to 2.48 comparing with Si waveguide. This work provides an insight that on-chip integration of graphene with CMOS-compatible silicon platform enables the realization of devices that possess many novel all-optical functions at telecommunication wavelength.

Table 1: The parameters used for numerical simulation in NLSE.

| Platform | Input pulse Width (fs) | Waveguide Length (mm) | Input peak Power (W) | $\beta_{TPA}$ (cm/GW) | FOM |
|---|---|---|---|---|---|
| G/Si hybrid (this work) | 1000 | 3.5 | 20 | 0.50 | 2.48 |
| Si (bulk) [18] | 130 | - | 4500 | - | 0.37 |
| Si [6] | 376 | 71 | 60 | 0.45 | 0.83 |
| Si-organic hybrid [42] | 1000 | 6.9 | 20 | 0.754 | 2.1 |
| Si$_{0.3}$Ge$_{0.7}$ [43] | 120 | 6 | 167 | 5.53 | 0.26 |
| α-Si [44] | 1600 | 10 | 4 | 0.62 | 2.0 |
| α-Si:H [45] | 180 | 7 | $2.8\times10^5$ | 4.1 | 0.66 |

**Acknowledgement**

The authors acknowledge the graphene transfer process by Dr. F.G. Yan from Institute of Semiconductors, Chinese Academy of Sciences in China. Financial support was provided by the National Natural Science Foundation of China (Grants 11504415, 11434041, 11574356 and 161635011), the Ministry of Science and Technology (MOST) of Peoples Republic of China (2016YFA0300600 and 2016YFA0301700), and the Key Research Program of Frontier Sciences, CAS (Grant No. QYZDB-SSW-JSC009).